\documentclass[]{mn2e}
\usepackage{mathrsfs,amssymb}
\usepackage{mathtools}
\usepackage{amsmath}
\usepackage{graphicx}
\usepackage{tabularx}
\usepackage{epsfig}
\usepackage{rotating}
\usepackage{pdflscape}
\usepackage{bm}
\newcommand       \Angstrom     {\,{\rm \AA}}

\newcommand       \cm           {\,{\rm cm}}

\newcommand       \kpc          {\,{\rm kpc}}

\newcommand       \mum          {\,{\rm \mu m}}

\newcommand       \simali       {\sim\,}
\newcommand       \magni    {\,{\rm mag}}
\newcommand     \gtsim  {\lower.5ex\hbox{$\buildrel > \over \sim$}}
\newcommand     \ltsim  {\lower.5ex\hbox{$\buildrel < \over \sim$}}
\newcommand     \simgt  {\lower.5ex\hbox{$\buildrel > \over \sim$}}
\newcommand     \simlt  {\lower.5ex\hbox{$\buildrel < \over \sim$}}
\newcommand       \Hatom    {\,{\rm H}}
\newcommand  \EWEBV  {{\rm EW}/E(B-V)}
\newcommand  \EWAV  {{\rm EW}/A_V}
\newcommand  \EWNH  {{\rm EW}/N_{\rm H}}
\newcommand       \RV           {{R_V}}
\newcommand       \AV           {{A_V}}

\newcommand       \EBV          {{E({\rm B-V})}}

\newcommand       \NH      {N_{\rm H}}

\def    \beq    {\begin{equation}}
\def    \eeq    {\end{equation}}
\def    \beqa   {\begin{eqnarray}}
\def    \eeqa   {\end{eqnarray}}
\title[The Missing Link between DIBs and $R_V$]{
Probing the Missing Link 
between the Diffuse Interstellar Bands and
the Total-to-Selective Extinction Ratio $R_V$
-- I. Extinction versus Reddening
}
\author[Li, Li \& Xiang]
            {Kaijun Li$^{1,2}$, %\thanks{E-mail: li$_{-}$kaijun@qq.com},
             Aigen Li$^{2}$\thanks{E-mail: lia@missouri.edu}, and
             F.Y.~Xiang$^{1,2}$\thanks{E-mail: fyxiang@xtu.edu.cn}\\
$^1$ Department of Physics, Xiangtan University, 
                  411105 Xiangtan, Hunan Province, China\\
$^2$ Department of Physics and Astronomy,
                  University of Missouri,
                  Columbia, MO 65211, USA\\
}

\begin{document}
\date{Received date  / Accepted date }
\pagerange{\pageref{firstpage}--\pageref{lastpage}} \pubyear{2019}

\maketitle

\label{firstpage}
\begin{abstract}
The carriers of the still (mostly) unidentified 
diffuse interstellar bands (DIBs) 
have been a long-standing mystery 
ever since their first discovery
exactly 100 years ago.
In recent years, the ubiquitous detection
of a large number of DIBs in a wide range
of Galactic and extragalactic environments
has led to renewed interest in connecting 
the occurrence and properties of DIBs
to the physical and chemical conditions of 
the interstellar clouds, %where DIBs are present.
%
%Of particular interest is whether and how 
%the carriers of DIBs respond to the ultraviolet (UV)
%starlight radiation which could potentially
%destroy the DIB carriers through photodissociation or
%generate more DIB carriers through photoionization
%if they are some kind of molecular ions.
%
%There is also renewed interest in tying
%the DIB carriers to the carriers of 
%the 2175$\Angstrom$ extinction bump
%and the far-UV extinction rise
%through a common origin or a similar 
%formation mechanism.
%
with particular attention paid to whether the DIB strength
is related to the shape of the interstellar extinction curve. 
To shed light on the nature and origin 
of the DIB carriers, we investigate 
the relation between the DIB strength and $R_V$, 
the total-to-selective extinction ratio,
which characterizes how the extinction varies 
with wavelength (i.e., the shape of the extinction curve). 
%and measures the quantities of the carriers
%of the 2175$\Angstrom$ bump
%and the far-UV extinction rise
%as well as the extent to which the interstellar
%UV radiation would be attenuated by interstellar dust.
We find that the DIB strength and $R_V$ are not related
if we represent the strength of a DIB 
by its reddening-normalized equivalent width (EW),
in contrast to the earlier finding of 
an anticorrelation in which the DIB strength 
is measured by the extinction-normalized EW.
This raises a fundamental question 
about the appropriate normalization 
for the DIB EW. We argue that
the hydrogen column density is 
a more appropriate normalization
than extinction and reddening. 
%
%
%We argue that, in addition to the possibility that
%DIBs and the UV extinction may have a common 
%origin or even share the same carriers,
%the UV radiation has complex effects
%of interplay between photodestruction 
%and photoionization on the DIB carriers 
%and thus there is no simple relation 
%between the DIB strength and the shape of
%the interstellar UV extinction curve.
%This is further complicated by the fact
%that $R_V$ may not be a viable indicator
%of the UV radiation.
%
\end{abstract}

\begin{keywords}
ISM: dust, extinction --- ISM: lines and bands
           --- ISM: molecules
\end{keywords}

\section{Introduction}\label{sec:intro}
%
%Since their discovery (Heger 1922), 
%the diffuse interstellar bands (DIBs) 
%have been a persistent spectroscopic 
%enigma -- representing unidentified 
%ingredients in the physics and chemistry 
%of the interstellar medium (ISM). The DIBs 
%are ubiquitous in the ISM--with the strongest 
%clearly visible toward even lightly reddened stars
%--though their relative strengths can vary in 
%different sight lines (e.g., Kre{\l}owski \& Walker 1987). 
%
%So far the vast majority of the DIBs 
%remain unidentified and
%the origin of the DIBs still remains enigmatic 
%(see Cami \& Cox 2014).
%
The enigmatic ``diffuse interstellar bands'' (DIBs)
are a set of over 600 non-stellar absorption features 
observed in starlight crossing interstellar clouds, 
spanning the wavelength range of 
the near ultraviolet (UV) 
at $\lambda\simgt 4000\Angstrom$ 
to the near infrared (IR) 
at $\lambda\simlt 1.8\mum$	
(e.g., see Sarre 2006).
It is frustrating that, 
except five DIBs were recently
identified as due to C$_{60}^{+}$,\footnote{%
   Campbell et al.\ (2015, 2016) and Walker et al.\ (2015) 
   measured the gas-phase spectrum of C$_{60}^{+}$ 
   at ultra-low temperatures and found that the spectral 
   characteristics of gas-phase C$_{60}^{+}$ 
   are in agreement with five DIBs at 9348.4, 9365.2,
   9427.8, 9577.0, and 9632.1$\Angstrom$
   (Cordiner et al.\ 2019),
   supporting the earlier assignment of 
   the 9577 and 9632$\Angstrom$ DIBs 
   to C$_{60}^{+}$ (Foing \& Ehrenfreund 1994) 
   which was based on the absorption spectrum
   recorded in a neon matrix (Fulara et al.\ 1993).
   Omont (2016) further proposed that fullerenes 
   of various sizes with endohedral or exohedral
   inclusions and heterofullerenes could be viable
   DIB carrier candidates.
   }
%However, the identification of C$_{60}^{+}$
%as a carrier of DIBs has not yet been proven 
%beyond reasonable doubt. 
%Galazutdinov et al.\ (2017) argued that,
%after correcting the contamination caused by
%the stellar Mg\,{\sc ii} line for the 9632$\Angstrom$ DIB,
%the relative strengths of the 9577 and 9632$\Angstrom$
%DIBs are too poorly correlated to be caused by 
%a single carrier (i.e., ${{\rm{C}}}_{60}^{+}$).
%More importantly, the 9428$\Angstrom$ absorption
%band which is more prominent than 
%the 9577 and 9632$\Angstrom$ bands
%in the experimental gas-phase spectrum 
%of ${{\rm{C}}}_{60}^{+}$ 
%is not seen in the lines of sight 
%where the 9577 and 9632$\Angstrom$ DIBs 
%are strong (see Galazutdinov \& Kre{\l}owski 2017,
%Cordiner et al.\ 2017).
%
% 
the vast majority of DIBs
have not yet been firmly identified,
despite an extensive observational, 
experimental, theoretical and computational 
exploration of an entire century 
--- it has been exactly 100 years 
since the first serendipitous detection 
of two DIBs (at 5780 and 5797$\Angstrom$) 
by Mary Lea Heger, 
then a graduate student at Lick Observatory
(see Heger 1922),
although their interstellar origin was not firmly
established until 15 years later by Merrill (1934).

Nevertheless, in recent years much progress 
has been made in detecting a large number of DIBs
in a wide variety of Galactic interstellar environments,
ranging from sightlines toward diffuse clouds,
translucent clouds, molecular clouds,
bright reflection nebulae, 
giant H\,{\sc ii} regions,
massive star forming regions,
massive young stellar clusters,
and the Galactic center
(see Snow \& McCall 2006,
Hobbs et al.\ 2008, 2009, 
Cox 2011, Cami \& Cox 2014, 
Kre{\l}owski 2018,
Sonnentrucker et al.\ 2018).
The detection of DIBs in extragalactic
environments has also been reported
(see Cordiner 2014),
ranging from the Local Group galaxies
including the Large and Small Magellanic Clouds 
(Ehrenfreund et al.\ 2002; Cox et al.\ 2006, 2007;
Welty et al.\ 2006; van Loon et al.\ 2013), 
M31 (Cordiner et al.\ 2008b, 2011),
M33 (Cordiner et al.\ 2008a),
M\,82 (Welty et al.\ 2014),
and the Antennae galaxies 
(Monreal-Ibero et al.\ 2018),
to more distant starburst galaxies 
(Heckman \& Lehnert 2000),
host galaxies of Type Ia supernovae 
(Sollerman et al.\ 2005, Cox \& Patat 2008,
Phillips et al.\ 2013),
interacting spiral galaxies (Monreal-Ibero et al.\ 2015),
intervening absorption systems toward quasars 
(Ellison et al.\ 2008) and damped Ly$\alpha$ absorbers 
at cosmological distances (DLAs; Junkkarinen et al.\ 2004, 
York et al.\ 2006, Lawton et al.\ 2008).
%
%Most notably, progress has also been made 
%in identifying several DIBs.
% 

Motivated by the ubiquitous detection of DIBs
in various Galactic and extragalactic environments,
in recent years much effort has been devoted
to exploring whether and how the presence,
strengths, and spectral profiles of DIBs are
affected by the physical and chemical conditions
of the interstellar environments
(e.g., see  Ruiterkamp et al.\ 2005,
Cox et al.\ 2006, Cox \& Spaans 2006,
Vos et al.\ 2011, Xiang et al.\ 2011,
Kos \& Zwitter 2013,
Clayton 2014, Sonnentrucker 2014).
There is a rich literature on DIB profiles 
and their variations in different lines of sight. 
A number of the DIBs (including those at
5780, 5797, and 6284$\Angstrom$) 
seen toward stars in the Orion Trapezium region 
are both broadened and shifted to the red
(Porceddu et al.\ 1992).
Blueshifted DIBs
(e.g., those at 5797 and 6614$\Angstrom$) 
which are broader than usual and whose
red wings are more prominent than usual
are seen toward the runaway star HD\,34078
(Galazutdinov et al.\ 2006)
which is currently interacting with 
a diffuse molecular cloud
(Boiss\'e et al.\ 2009).
Anomalously broad DIBs 
at 5780, 5797, 6196, and 6614$\Angstrom$ 
with remarkable extended tails toward red
are found in absorption along the line of sight 
to Herschel 36, an O star multiple system
%and a member of the young star cluster NGC\,6530
illuminating the bright Hourglass nebula
of the H\,{\sc ii} region Messier\,8
(Dahlstrom et al.\ 2013, York et al.\ 2014).

In addition to the environmental variations 
of the DIB spectral profiles, the strength of
a DIB, as measured by its equivalent width (EW),
is also known to vary significantly from 
sightline to sightline. This variation appears 
to depend on the local environmental conditions 
(see Cami \& Cox 2014 and references therein).\footnote{%
  For example, DIBs were observed to be weaker 
  by factors of $\simali$2 or more on a per unit reddening
  basis in photon-dominated regions 
  (PDR; see Jenniskens et al.\ 1994). 
  }
It has long been known that the DIB strength 
correlates with the interstellar reddening
$E(B-V)$ caused by solid dust particles\footnote{%
   Also known as the ``color excess'',
   the interstellar reddening is defined 
   as the extinction difference between
   two wavebands, e.g., $E(B-V) = A_B-A_V$
   is the difference between the $B$-band
   extinction ($A_B$) at $\lambda\approx4400\Angstrom$
   and that of the $V$-band  ($A_V$) 
   at $\lambda\approx5500\Angstrom$.
   }
--- as a matter of fact, this correlation was
one of the principal arguments supporting
an interstellar origin of DIBs (Merrill 1934).
Although the correlation between the DIB strength
and the interstellar reddening does not necessarily
mean that the interstellar reddening and DIBs
share a common carrier --- it may merely reflect
the fact that the DIB carriers are well mixed with
dust and gas in interstellar clouds so that 
both the amount of DIB carriers and 
the quantity of interstellar dust grains 
are linearly proportional to the amount
of interstellar gas along the line of sight.
Nevertheless, the DIB carriers and interstellar 
dust grains could be physically related, 
e.g., the latter could protect the former 
from photodestruction by shielding 
the former from the UV starlight. 

In this context, it would be of great value 
to explore how the DIB strength varies
with $R_V\equiv A_V/E(B-V)$,
the total-to-selective extinction ratio, 
which characterizes the steepness of
the UV extinction: for lines of sight with
a smaller $R_V$, the UV extinction often
increases more steeply with $\lambda^{-1}$,
the inverse wavelength (e.g., see Figures~4,6
of Cardelli, Clayton \& Mathis 1989, 
hereafter CCM). On a per unit $A_V$ basis,
a smaller $R_V$ implies a more severe 
attenuation of the UV radiation and thus
a more effective protection of the DIB carriers.
While this scenario is complicated by
the fact that those lines of sight 
with a smaller $R_V$ often subject to 
a smaller amount of visual extinction
($A_V$), observationally, this has been 
demonstrated in the $\sigma$ Sco cloud 
toward HD\,147165 
(for which $R_V\approx4.25$, Lewis et al.\ 2005) 
and the $\zeta$ Oph cloud toward HD\,149757 
(for which $R_V\approx3.08$, Fitzpatrick \& Massa 2007).
While the interstellar reddening 
[$E(B-V)\approx0.34$ for $\sigma$ Sco
and $E(B-V)\approx0.32$ for $\zeta$ Oph]
is similar for these two clouds,
the 5780$\Angstrom$ DIB of $\sigma$ Sco
is substantially stronger than that of $\zeta$ Oph.
However, it is worth noting that the strengths of 
the 5797$\Angstrom$ DIB of these very same 
two clouds are nearly identical
(Kre{\l}owski \& Westerlund 1988).
The effects of the UV starlight on
the DIB strength has also been 
observationally studied by Cami et al.\ (1997) 
and Sonnentrucker et al.\ (1997)
for a larger number of DIBs 
and a larger sample of sightlines.
%for a larger number of (44) DIBs 
%and a larger sample of (14) sightlines.
%
%tracing different physical conditions. 

To shed light on the physical and chemical
nature of the still unidentified DIB carriers,
we have initiated a program to explore 
the possible relations between the DIB
strengths and the various interstellar parameters
(e.g., extinction, UV radiation, and gas densities).
In this work, we quantitatively examine 
how DIBs vary with $R_V$, first for the sightlines 
toward the H\,{\sc ii} region in M17 
and then for a larger sample 
(\S\ref{sec:DIBvsRV}).
It is found that the DIB strength, 
measured as $\EWEBV$, 
the EW of a DIB normalized by reddening, 
is not correlated with $R_V$.
This is in contrast to the earlier finding 
of an anticorrelation 
made by Ram\'irez-Tannus et al.\ (2018)
who normalized the DIB EW by $A_V$.
The results are discussed in \S\ref{sec:discussion}
and summarized in \S\ref{sec:summary}.
%This is in contrast to the earlier finding 
%of an anticorrelation 
%made by Ram\'irez-Tannus et al.\ (2018).
%who normalized the DIB strength by $A_V$.
%With $A_V = R_V\times E(B-V) \propto R_V$,
%an anticorrelation between $\EWAV$,
%the EWs of DIBs normalized by $A_V$,
%and $R_V$ is not unexpected (\S\ref{sec:discussion}).
%This raises a fundamental question 
%about the appropriate normalization 
%for the DIB EW. We argue that
%the hydrogen column density is 
%a more appropriate normalization factor
%than extinction and reddening. 
%In \S\ref{sec:summary} we summarize 
%the major results.

\section{Is the DIB Strength Related to $R_V$?}\label{sec:DIBvsRV}
\begin{figure*}
\centering
\begin{tabular}{c}
%\vspace{-1.0em}
\hspace{-1.5cm}
\includegraphics[width=1.2\textwidth]{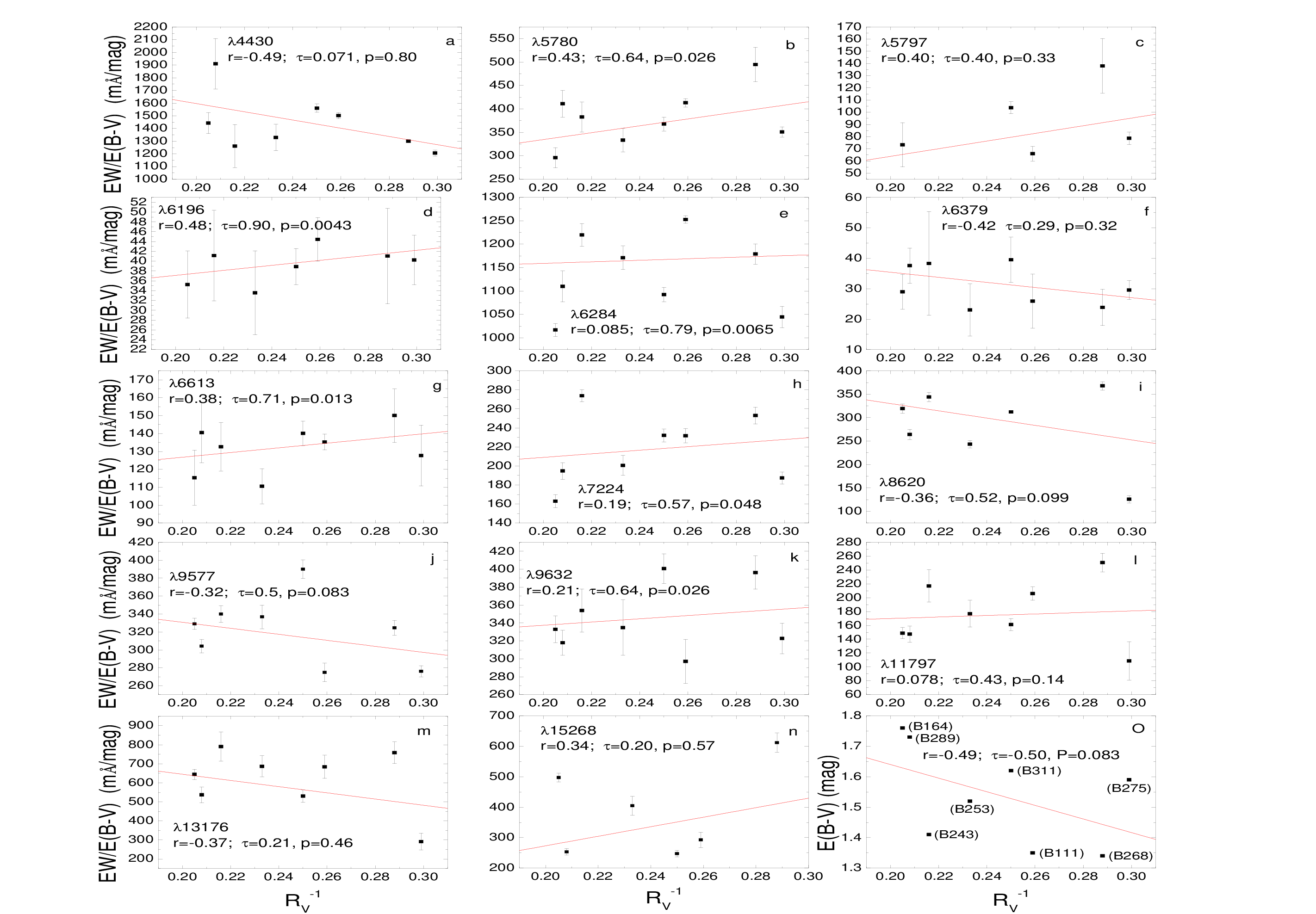}
\end{tabular}
\caption{\label{fig:EWvsEBV}
             Correlation of the reddening-normalized DIB EW
             with $R_V^{-1}$ for the lines of sight toward eight
             young OB stars in the giant H\,{\sc ii} region M17
             for 14 DIBs at 4430$\Angstrom$ (a), 5780$\Angstrom$ (b), 
             5797$\Angstrom$ (c), 6196$\Angstrom$ (d),
             6284$\Angstrom$ (e), 6379$\Angstrom$ (f), 
             6614$\Angstrom$ (g), 7224$\Angstrom$ (h), 
             8620$\Angstrom$ (i), 9577$\Angstrom$ (j), 
             9632$\Angstrom$ (k), 11797$\Angstrom$ (l), 
             13176$\Angstrom$ (m), and 15268$\Angstrom$ (n).
             Also shown is the correlation between the reddening
             $E(B-V)$ and $R_V^{-1}$ (o). 
             Labeled in each subfigure are 
             the Pearson correlation coefficient $r$, 
             the Kendall's $\tau$ coefficient 
             and the significance level $p$.
             The star identifiers are also 
             labeled in Figure~\ref{fig:EWvsEBV}(o).
             }
\end{figure*}

Following Ram\'irez-Tannus et al.\ (2018), 
we first consider the prominent DIBs seen 
in M17, a giant H\,{\sc ii} region. 
Located in the Carina-Sagittarius spiral arm
of the Galaxy at a distance of
$\simali$1.98$\kpc$, M17 is one of 
the brightest and best-studied giant H\,{\sc ii} region
(Hoffmeister et al.\ 2008, Povich et al.\ 2009, 
Ram\'irez-Tannus et al.\ 2017). 
M17 is selected for this study because the sightlines 
toward M17 exhibit a significant spread in 
both extinction ($A_V$\,$\sim$\,3--15$\magni$)
and $R_V$ ($\simali$2.8--5.5). 
Hanson et al.\ (1997) investigated the behavior 
of the DIBs along the sightlines toward M17, 
over such a wide extinction range.
They found that the DIBs show little change 
in spectral shape.
%
%Hanson et al.\ (1997):
% We have also used the M17 O stars to study the dust 
%properties in the local cloud and the behavior of 
%the DIBs along this sight line, over the extinction range 
%of $A_V$\,=\, 3--10. The DIBs over this extinction range 
%show little change in spectral shape 
% nor a significant increase 
%in strength. We suggest the features are already saturated 
%at small AV, or the material local to M17, where the increased
%extinction is being traced, does not contain the carriers 
%of the DIB feature. 
%
Ram\'irez-Tannus et al.\ (2018) obtained 
the 300--2500\,nm spectra of 
11 pre-main sequence OB stars 
with the X-shooter Spectrograph mounted 
on the ESO Very Large Telescope (VLT).
They determined the reddening, visual extinction,
and $R_V$ for the lines of sight toward these stars.
They also measured the EWs of 14 prominent DIBs
for most of these sightlines.
As tabulated in Table~\ref{tab:dibdata},
we adopt the reddening $E(B-V)$, $R_V$, 
and DIB EW data of Ram\'irez-Tannus et al.\ (2018) 
and examine the correlation between the DIB EW
and $R_V^{-1}$ in M17.
For different lines of sight crossing different
amounts of interstellar matter, the DIB EW
is expected to correlate with $E(B-V)$.\footnote{%
  In the ISM, dust and gas are well mixed
  as indicated by the relatively constant
  gas-to-extinction ratio, 
  $N_{\rm H}/E(B-V)\approx 5.8\times 10^{21}\Hatom\cm^{-2}\magni^{-1}$
  (Bohlin et al.\ 1978).
  Therefore, any two interstellar quantities that depend on 
  either the amount of dust or the amount of gas 
  in the line of sight will tend to,
  to some extent, correlate with each other.
  To explore the correlation between the DIB strength
  and the extinction parameters,  
  the common dependence on the reddening
  $E(B-V)$ has to be cancelled out
  (e.g., see Witt et al.\ 1983, Xiang et al.\ 2011).
  } 
Therefore, to cancel out the common 
correlation between the DIB EW and $E(B-V)$
among the various lines of sight, 
we normalize the DIB EWs by $E(B-V)$.
We perform a correlation analysis 
between the reddening-normalized EWs 
of 14 DIBs, $\EWEBV$, and $R_V^{-1}$
for the lines of sight toward M17.
As shown in Figure~\ref{fig:EWvsEBV}, 
for all 14 DIBs at 4430, 5780, 5797, 6196,
6284, 6379, 6614, 7224, 8620, 9577, 
9632, 11797, 13176, and 15268$\Angstrom$,
the Pearson correlation coefficient $r$ 
never exceeds 0.50, 
indicating that $\EWEBV$ and $R_V^{-1}$
are not correlated.\footnote{%
   Although somewhat arbitrary, 
   we suggest that, for two variables
   to be considered to be (even weakly) correlated, 
   the Pearson correlation coefficient ($r$) 
   should at least exceed 0.5
   (e.g., see {\sf https://explorable.com/statistical-correlation}).
   }
This is also supported by the Kendall's $\tau$ test
(see Figure~\ref{fig:EWvsEBV}). 

\begin{figure*}
\centering
\begin{tabular}{c}
%\vspace{-1.0em}
\includegraphics[width=1.2\textwidth]{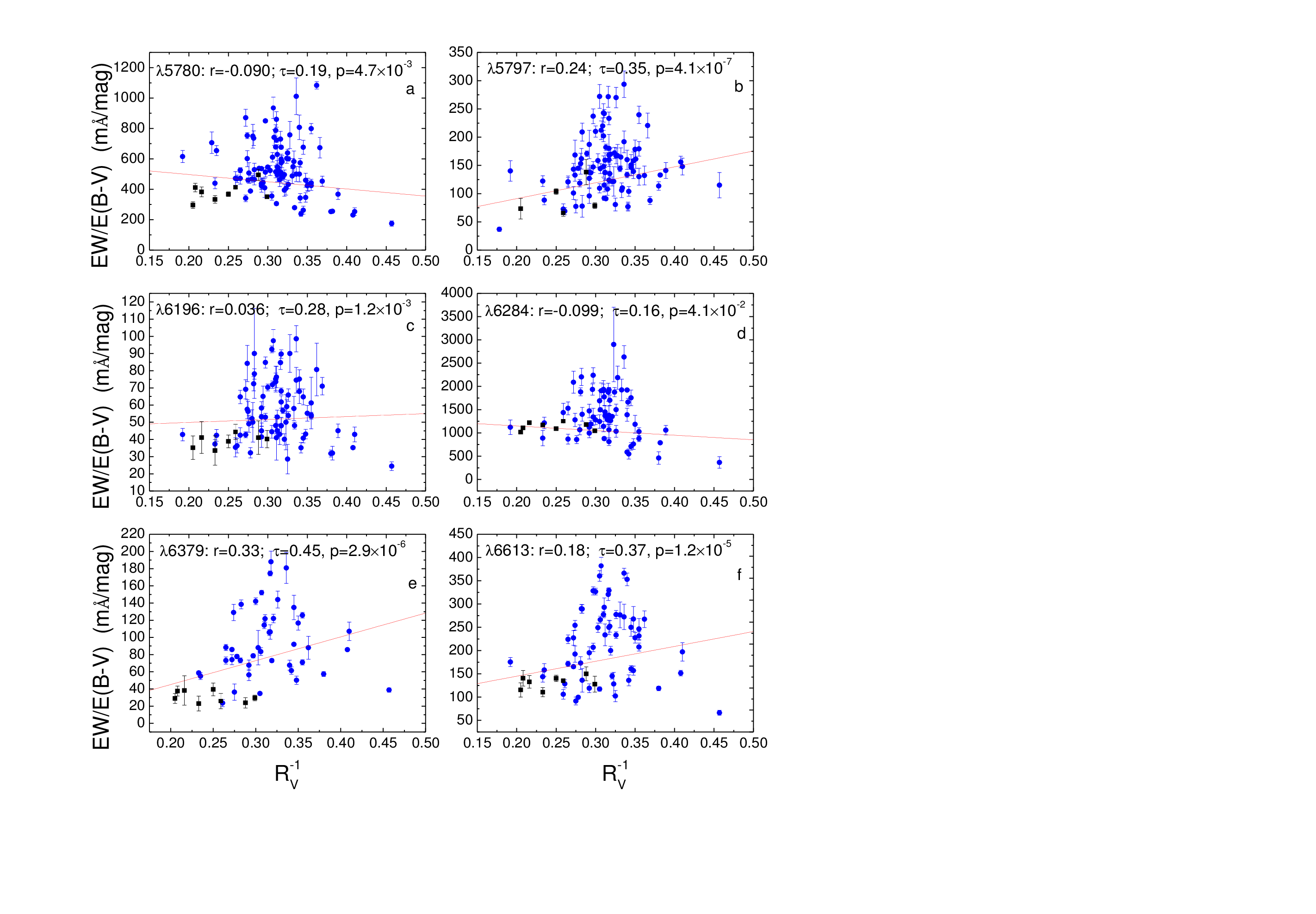}
\end{tabular}
\caption{\label{fig:EW2EBVXiang}
             Correlation of the reddening-normalized DIB EW
             with $R_V^{-1}$ for six DIBs at 5780$\Angstrom$ (a), 
             5797$\Angstrom$ (b), 6196$\Angstrom$ (c),
             6284$\Angstrom$ (d), 6379$\Angstrom$ (e), 
             and 6614$\Angstrom$ (f) for a large sample
             of 97 sightlines of which both EWs and $R_V$ 
             have been compiled from the literature 
             by Xiang et al.\ (2017).
             }  
\end{figure*}

We have also performed the Pearson correlation 
analysis and the Kendall's $\tau$ test
for six of these 14 DIBs for a large sample
of 97 sightlines of which both EWs and $R_V$ 
have been compiled from the literature 
by Xiang et al.\ (2017).
As illustrated in Figure~\ref{fig:EW2EBVXiang},
no correlation is found between
$\EWEBV$ and $R_V^{-1}$.

\section{Discussion}\label{sec:discussion}
\begin{figure*}
\centering
\begin{tabular}{c}
%\vspace{-1.0em}
\hspace{-1.5cm}
\includegraphics[width=1.2\textwidth]{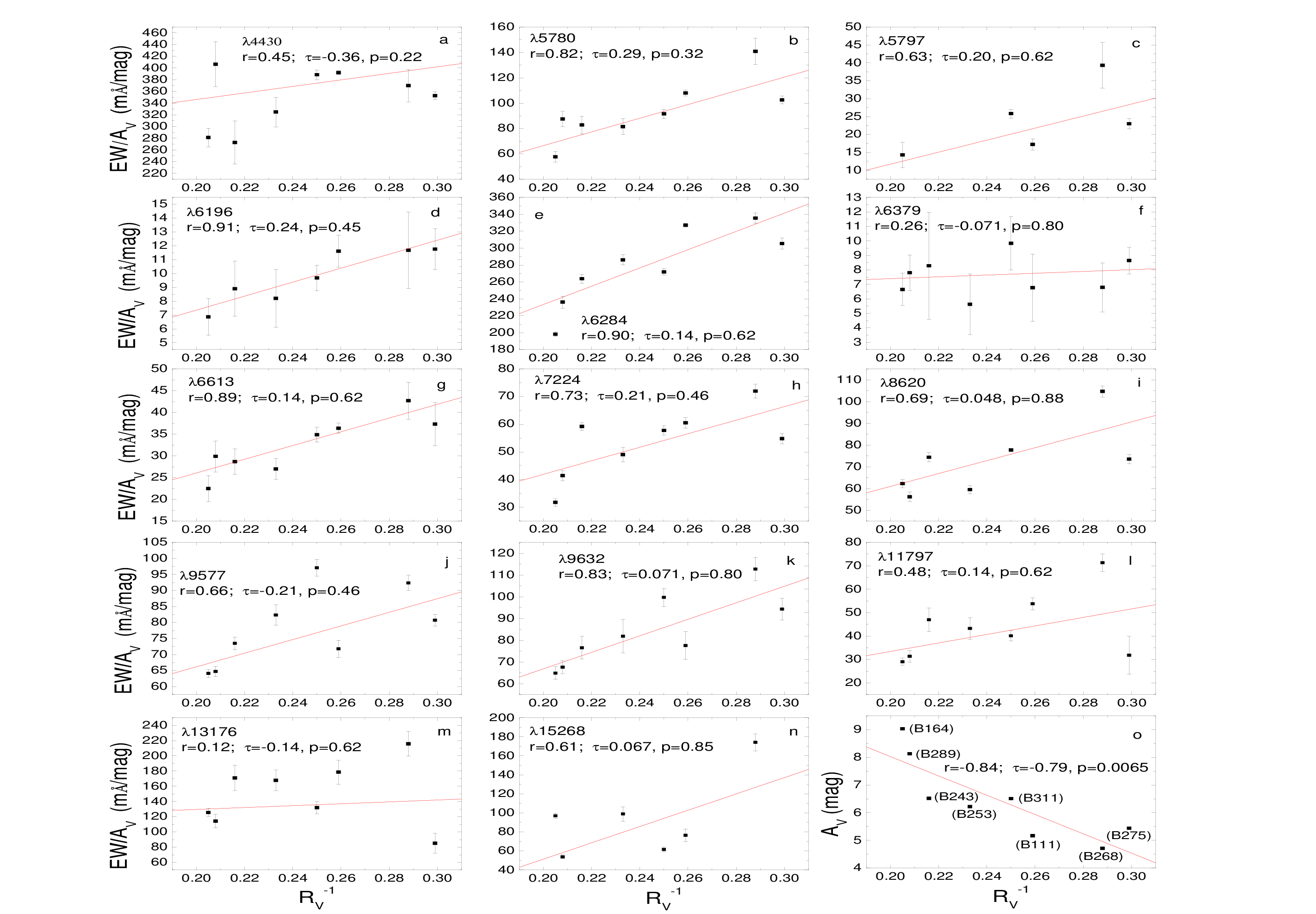}
\end{tabular}
\caption{\label{fig:EWvsAV}
             Same as Figure~\ref{fig:EWvsEBV}
             but for $\EWAV$, the extinction-normalized DIB EWs.
             Figure~\ref{fig:EWvsAV}(o) shows the correlation 
             between $A_V$ and $R_V^{-1}$. 
             }  
\end{figure*}

%
%%% York, D. 1971: Structure in Extinction Curve %%%
%
%\subsection{$\EWEBV$ vs. $\EWAV$}\label{sec:EBVorAV}
%                    or to normalize the DIB Strength 
%                    by $E(B-V)$ or by $A_V$: that is 
%                    the question!\label{sec:EBVorAV}}
%
Ram\'irez-Tannus et al.\ (2018) investigated the correlation
between $\EWAV$, the extinction-normalized DIB EWs,
and $R_V^{-1}$ for the 14 prominent DIBs seen in M17. 
As reproduced here in Figure~\ref{fig:EWvsAV}, 
it is apparent that, with the Pearson correlation 
coefficient $r$ exceeding 0.80 for five of the 14 DIBs 
and exceeding 0.60 for 10 of the 14 DIBs, 
$\EWAV$ appears to correlate with $R_V^{-1}$
for the vast majority of the DIBs seen in M17.
This is in stark contrast to our finding that 
there seems to be no correlation between 
$\EWEBV$ and $R_V^{-1}$ (see \S\ref{sec:DIBvsRV}).

The major difference between our approach 
and that of Ram\'irez-Tannus et al.\ (2018) 
lies in the normalization: while we normalize
the DIB EW by reddening $\EBV$,  
Ram\'irez-Tannus et al.\ (2018) took the visual
extinction $\AV$ as the normalization.
We argue that the anticorrelation 
between $\EWAV$ and $R_V^{-1}$ 
may be related to the fact that,
for the M17 sample of Ram\'irez-Tannus et al.\ (2018), 
$A_V$ and $R_V$ are themselves correlated.
%$A_V = E(B-V)\times R_V\propto R_V$.
As shown in Figure~\ref{fig:EWvsAV}(o),
with a Pearson correlation coefficient of $r\approx-0.84$
and a Kendall correlation coefficient of $\tau\approx-0.79$ 
and a corresponding probability $p\approx0.0065$ 
of a chance correlation, an anticorrelation between $A_V$
and $R_V^{-1}$ is apparent.\footnote{%
   Ram\'irez-Tannus et al.\ (2018) determined 
   $A_V$ and $E(B-V)$ by comparing the observed 
   magnitudes of the target stars to the intrinsic 
   ones of the same spectral type, i.e., $A_V= V-V_0$,
   and $E(B-V) = (B-V)-(B-V)_0$.
   They derived $R_V$ using the relation reported
   in Fitzpatrick \& Massa (2007):
   $R_V \approx -0.26 + 1.19\,(A_V-A_K)/(A_B-A_V)$,
   where $A_K$ is the $K$-band extinction.
   This explains why the Pearson correlation coefficient
   $r$ between $A_V$ and $R_V^{-1}$ is not $-1$
   which should have been the case if $R_V$ was derived
   directly from $A_V/E(B-V)$.
   }
Therefore, even if DIBs do not correlate with $R_V$ at all, 
the intrinsic anticorrelation between $A_V$ and $R_V^{-1}$
would lead to $\EWAV$ to correlate with $R_V^{-1}$.
On the other hand, as demonstrated 
in Figure~\ref{fig:EWvsEBV}(o), $E(B-V)$ is
not related to $R_V^{-1}$.
%

%%%% Below: Revision %%%%
This raises a fundamental question:
when one explores the possible relations 
between DIBs and other interstellar parameters 
or among different DIBs, what is a more
appropriate normalization, $\AV$ or $\EBV$?
%Note that the whole reason for normalizing
%the DIB EW by $\AV$ or $\EBV$ is to eliminate 
%the common correlation among the column 
%densities of the DIB carriers, dust and gas.
%
%
At a first glance, $\AV$ appears to be a better
normalization since $\AV$ is a direct tracer of 
the dust column density.
However, $\EBV$ is a better discriminator 
of dust size and therefore of $\RV$
(e.g., see Figures~22.7, 22.8 of Draine 2011)
in the sense that larger grains intend 
to be ``grayer'' and have larger $R_V$.
%and result in a smaller reddening
%(on a per unit dust mass basis). 
When correlating the DIB EW with $\RV$,
it thus seems more appropriate to normalize
the DIB EW with $\EBV$ than $\AV$.
In this way, any intrinsic relation 
between $E(B-V)$ and $R_V$ 
would have been cancelled out. 
%
%While $\EBV$ has been used frequently in the past 
%as a stand-in for $\Ndust$, this is merely 
%a matter of convenience.\footnote{% 
%   {\bf Observationally, it is easier to measure $\EBV$
%   than $\AV$. For a given sightline, one can derive 
%   $\EBV$ from the spectral types and UBV photometry data 
%   of background stars. To obtain $\AV$ in the absence of 
%   accurate trigonometric parallaxes, one requires 
%   both $\EBV$ and $\RV$, while the latter is much 
%   more difficult to determine. 
%   Consequently, $\AV$ is known for far fewer stars
%   than $\EBV$.
%   }
%
%
%
%For this reason, published values of E(B-V) exist for tens of
%thousands and soon for millions of stars. To obtain values for A_v 
%
%The color excess E(B-V) = A_v/R_v is a biased tracer of the dust
%column density in the sense that it preferentially tracks dust with
%low values of R_v. 
%
%It is E(B-V) that has an implicit inverse proportional relationship
%with R_v. 
%
%As it turns out, DIBs carriers in the free state also are similarly
%biased toward dust with low values of R_v. 
%
%As a result, the ratio EW(DIB)/E(B-V) is largely independent of R_v, 
%
%For this very reason, we also find consistently that EW(DIB) 
%exhibits stronger correlations with E(B-V) 
%compared to correlations with A_v.
%
%
Also, it is well known and can be easily verified 
with the DIB EW data available in the public domain\footnote{%
   See, e.g., the University of Chicago DIB database at
   {\sf http://dib.uchicago.edu/public/index.html}.
   }
that the DIB strengths for most of the strong DIBs 
(e.g., those at 4430, 5780, 5797, 6284, 6613$\Angstrom$)
are more strongly correlated with $E(B-V)$ than with $A_V$. 
This appears to support $\EBV$ as a more favorable
normalization than $\AV$.
On the other hand, this could also be considered
as evidence for supporting that DIBs actually 
physically anticorrelate with $\RV$,
at least through the so-called 
``skin'' or ``edge'' effect (Snow \& Cohen 1974),
i.e., in dense molecular clouds characterized by 
larger-than-average $R_V$ values, DIBs are weak 
or even completely absent at the cloud cores 
but grow in strength toward the cloud edges.
This is possibly caused by the accretion of 
the DIB carriers onto the surfaces of large dust grains 
under conditions which favor dust growth 
in dense molecular clouds. As a result,
DIB carriers are relatively under-abundant 
and thus many DIBs exhibit smaller strengths 
in these environments, i.e., places where dust 
with larger values of $R_V$ typically resides.\footnote{%
  Indeed, Hansen et al.\ (1997) have already noted 
  that the DIBs observed in the direction of M17, 
  over the extinction range of $A_V$\,=\,3--10$\magni$, 
  does not show any significant increase in strength. 
  They suggested that either the DIB features are already 
  saturated at a small value of $A_V$,
  or that the interstellar material local to M17, 
  where the increased extinction is being traced, 
  does not contain DIB carriers. 
  }

We argue that neither $\AV$ nor $\EBV$ is an accurate
tracer of the dust column density since both quantities
involve the properties (e.g., size, composition) of the dust 
along the line of sight which exhibit regional variations.
We suggest the hydrogen column density ($\NH$) is 
a more appropriate normalization
than $\AV$ and $\EBV$ since $\NH$ directly measures
the amount of interstellar material along a given sightline,
while both $\AV$ and $\EBV$ are actually only used
as proxies for $\NH$.
Unfortunately, in the literature there is no $\NH$ information 
for the M17 sightlines of interest here. 
This prevents us from a quantitative analysis 
of the relation between $\EWNH$ and $\RV^{-1}$.

\section{Summary}\label{sec:summary}
%
%To explore whether and how the mysterious DIBs 
%would be related to the interstellar UV extinction 
%and respond to the UV radiation field and thus to 
%gain insight into the unidentified identity of 
%the DIB carriers, 
%
We have examined the relation 
between the DIB strength and $R_V$
which characterizes how the extinction 
varies with wavelength,
% and measures 
%the degree how severe the interstellar UV
%radiation would be attenuated by interstellar dust,
first for 14 DIBs in eight lines of sight toward
young OB stars in the giant H\,{\sc ii} region M17
and then for six of these 14 DIBs in a large sample
of 97 lines of sight compiled from the literature. 
It is found that the DIB strength, measured as
the reddening-normalized DIB EW, is not correlated
with $R_V$, in contrast to the earlier finding of 
an anticorrelation between the extinction-normalized 
DIB EW and $R_V$.
We argue that, when comparing the DIB EW 
with $\RV$, neither $\AV$ nor $\EBV$ is an ideal 
normalization since $\AV$ is usually intrinsically 
higher for regions with larger $\RV$ 
(i.e., $\AV\propto\RV$)
while $\EBV$ preferably probes
the surface layers of dense molecular cloud cores.
We suggest that the really appropriate normalisation 
for the DIB EW, on physical grounds, 
would be $\NH$, the hydrogen column density. 
%
%

%%%%% Acknowledgments %%%%%%
\section*{Acknowledgements}
We thank the anonymous referee
for his/her very helpful comments and
suggestions which have considerably 
improved the presentation of this work.
KJL and FYX are supported in part by 
the Joint Research Funds in Astronomy
(U1731106, U1731107 and U1531108) 
under cooperative agreement between 
the National Natural Science Foundation of China 
and Chinese Academy of Sciences.
AL is supported in part by NSF AST-1816411.

%%%%%%%%%%%%%%% References %%%%%%%%%%%%%
%\bibliographystyle{mn2e}

%%%% Figure 1: EW/E(B-V) vs. 1/R_V  %%%%%%
%%%%%% Figure 1 %%%%%%%%

%%%% Figure 2: EW/E(B-V) vs. 1/R_V for all sources %%%%%%
%%%%%% Figure 2 %%%%%%%%

%%%% Figure 3: EW/A_V vs. 1/R_V  %%%%%%
%%%%%% Figure 3  %%%%%%%%

%%%% Figure 4: EW/A_V vs. 1/R_V for all sources %%%%%%
%\begin{figure*}
%\centering
%\begin{tabular}{c}
%\vspace{-1.0em}
%\includegraphics[width=1.2\textwidth]{EW2AVXiang.eps}
%
%\end{tabular}
%\caption{\footnotesize
%             \label{fig:EW2AVXiang}
%             $W_{\rm DIB}$/$A_V$ vs. $1/R_V$
%             for all sources.
%             }  
%\end{figure*}
%%%%%% Figure 4 %%%%%%%%

%%%%%%%%%%%%%%%%%%%%% Table 1 %%%%%%%%%%%%%%%%
\begin{table*}
%\scriptsize
\begin{center}
\caption{\label{tab:dibdata}
               Extinction and DIB Properties
               in the Lines of Sight toward 
               Eight Young OB Stars
               (B111, B164, B243, B253, B268,
               B275, B289 and B311) in M17
               (Ram\'irez-Tannus et al.\ 2018).
               }
\begin{tabular}{lcccccccr}
\hline
Extinction and 
& B111 & B164 & B243 & B253 
& B268 & B275 &B289 & B311\\ 
DIB Properties & & & & & & & &\\
\hline
$R_{V}$                     &  3.86          &  4.87             &    4.64                &   4.29                  &    3.47                &  3.35              &   4.80              &  4.00              \\
$A_{V}$/mag                     &  5.17          &  9.03             &    6.52                &   6.22                  &    4.71                &  5.44              &   8.13              &  6.51              \\
$E(B-V)$/mag                     &  1.35          &  1.76             &    1.41                &   1.52                  &    1.34                &  1.59              &   1.73              &  1.62              \\  
${\rm EW}(4430)/{\rm m\AA}$   &  2027$\pm$32   &  2539$\pm$145     &     1779$\pm$240       &    2020$\pm$158         &     1742$\pm$130       &   1919$\pm$36      &    3304$\pm$311     &   2528$\pm$53      \\  
${\rm EW}(5780)/{\rm m\AA}$   &   558$\pm$12   &  521$\pm$38       &     540$\pm$45         &    507$\pm$38           &     663$\pm$49         &   558$\pm$17       &    711$\pm$49       &   596$\pm$24       \\  
${\rm EW}(5797)/{\rm m\AA}$   &    89$\pm$8    &  129$\pm$32       &     --                 &    --                   &     185$\pm$30         &   125$\pm$8        &    --               &   168$\pm$8        \\  
${\rm EW}(6196)/{\rm m\AA}$   &    60$\pm$6    &  62$\pm$12        &     58$\pm$13          &    51$\pm$13            &     55$\pm$13          &   64$\pm$8         &    --               &   63$\pm$6         \\  
${\rm EW}(6284)/{\rm m\AA}$   &  1691$\pm$11   &  1790$\pm$24      &     1720$\pm$34        &    1708$\pm$38          &     1580$\pm$29        &   1661$\pm$36      &    1920$\pm$57      &   1770$\pm$25      \\  
${\rm EW}(6379)/{\rm m\AA}$   &    35$\pm$12   &  51$\pm$10        &     54$\pm$24          &    35$\pm$13            &     32$\pm$8           &   47$\pm$5         &    65$\pm$10        &   64$\pm$12        \\  
${\rm EW}(6614)/{\rm m\AA}$   &   188$\pm$6    &  203$\pm$27       &     187$\pm$19         &    168$\pm$15           &     201$\pm$20         &   203$\pm$27       &    243$\pm$29       &   227$\pm$11       \\  
${\rm EW}(7224)/{\rm m\AA}$   &   313$\pm$10   &  287$\pm$12       &     386$\pm$9          &    305$\pm$16           &     339$\pm$12         &   298$\pm$10       &    337$\pm$15       &   376$\pm$11       \\  
${\rm EW}(8620)/{\rm m\AA}$   &  --            &  562$\pm$17       &     485$\pm$14         &    370$\pm$12           &     493$\pm$12         &   400$\pm$12       &    457$\pm$19       &   506$\pm$5        \\  
${\rm EW}(9577)/{\rm m\AA}$   &   371$\pm$14   &  579$\pm$11       &     479$\pm$13         &    512$\pm$20           &     435$\pm$11         &   439$\pm$10       &    526$\pm$13       &   632$\pm$17       \\  
${\rm EW}(9632)/{\rm m\AA}$   &   401$\pm$33   &  586$\pm$26       &     499$\pm$34         &    509$\pm$47           &     531$\pm$25         &   513$\pm$27       &    550$\pm$24       &   649$\pm$27       \\  
${\rm EW}(11797)/{\rm m\AA}$  &   278$\pm$13   &  262$\pm$14       &     306$\pm$33         &    269$\pm$29           &     336$\pm$18         &   173$\pm$44       &    255$\pm$20       &   261$\pm$14       \\  
${\rm EW}(13176)/{\rm m\AA}$  &   923$\pm$82   &  1134$\pm$47      &     1115$\pm$108       &    1044$\pm$84          &     1016$\pm$76        &   463$\pm$70       &    928$\pm$71       &   859$\pm$53       \\  
${\rm EW}(15268)/{\rm m\AA}$  &   395$\pm$34   &  875$\pm$25       &     --                 &    615$\pm$47           &     820$\pm$42         &   --               &    438$\pm$18       &   401$\pm$17       \\  
\hline
\end{tabular}
\end{center}
\end{table*}        

%\clearpage
        
\end{document}